\documentclass[preprint,3p,number]{elsarticle} % maquetación legible en una columna
\pdfoutput=1

\usepackage{graphicx}
\usepackage{amsmath,amssymb}
\usepackage{subfig}
\usepackage{booktabs}
\usepackage{multirow}
\usepackage{arydshln}
\usepackage{xcolor}
\usepackage{soul}
\usepackage{hyperref}

\journal{Knowledge-Based Systems}

\usepackage{graphicx}
\usepackage{amssymb}
\usepackage{pifont}
\newcommand{\cmark}{\checkmark}
\newcommand{\xmark}{\ding{55}}

\begin{document}

\begin{frontmatter}

\title{Difficulty-Aware Score Generation for Piano Sight-Reading}

% ==== Autores y afiliaciones ====
\author[aff1]{Pedro Ramoneda\corref{cor1}}
\ead{pedro.ramoneda@upf.edu}
\author[aff2]{Masahiro Suzuki}
\author[aff2]{Akira Maezawa}
\author[aff1]{Xavier Serra}

\address[aff1]{Music Technology Group, Universitat Pompeu Fabra, Barcelona, Spain}
\address[aff2]{Yamaha Corporation, Hamamatsu, Japan}

\cortext[cor1]{Corresponding author.}

% ==== Abstract ====
\begin{abstract}
Adapting learning materials to the level of skill of a student is important in education. In the context of music training, one essential ability is sight-reading---playing unfamiliar scores at first sight---which benefits from progressive and level-appropriate practice. However, creating exercises at the appropriate level of difficulty demands significant time and effort. We address this challenge as a controlled symbolic music generation task that aims to produce piano scores with a desired difficulty level. Controlling symbolic generation through conditioning is commonly done using control tokens, but these do not always have a clear impact on global properties, such as difficulty. To improve conditioning, we introduce an auxiliary optimization target for difficulty prediction that helps prevent conditioning collapse---a common issue in which models ignore control signals in the absence of explicit supervision. This auxiliary objective helps the model to learn internal representations aligned with the target difficulty, enabling more precise and adaptive score generation. Evaluation with automatic metrics and expert judgments shows better control of difficulty and potential educational value. Our approach represents a step toward personalized music education through the generation of difficulty-aware practice material.
\end{abstract}

\begin{keyword}
Music Generation \sep Automatic Composition \sep Education Technology
\end{keyword}

\end{frontmatter}

\section{Introduction}

Sight-reading is the ability to perform a piece of music at first sight without prior preparation, and it is a core skill in music education and a standard component of piano exams and curricula around the world~\cite{abrsm2024sightreading,trinity2021piano,yamaha}, with previous research focusing on their analysis and assessment~\cite{cheng2008nmf,huang2019sightreading,elliott1982sightreading}. Its development requires frequent exposure to unfamiliar and well-targeted material that matches the current abilities of the learner~\cite{farley2014pulse}. However, generating suitable exercises for sight-reading practice is a time-consuming task, and most students and teachers rely on static, non-personalized resources. As a result, learners often face materials that are not adapted to their needs, are too easy to play, or are too difficult to support meaningful progress~\cite{hayward2009relationships}.

\begin{figure}[t!]
    \centering
    \includegraphics[width=0.5\linewidth, trim=25pt 25pt 25pt 25pt, clip]{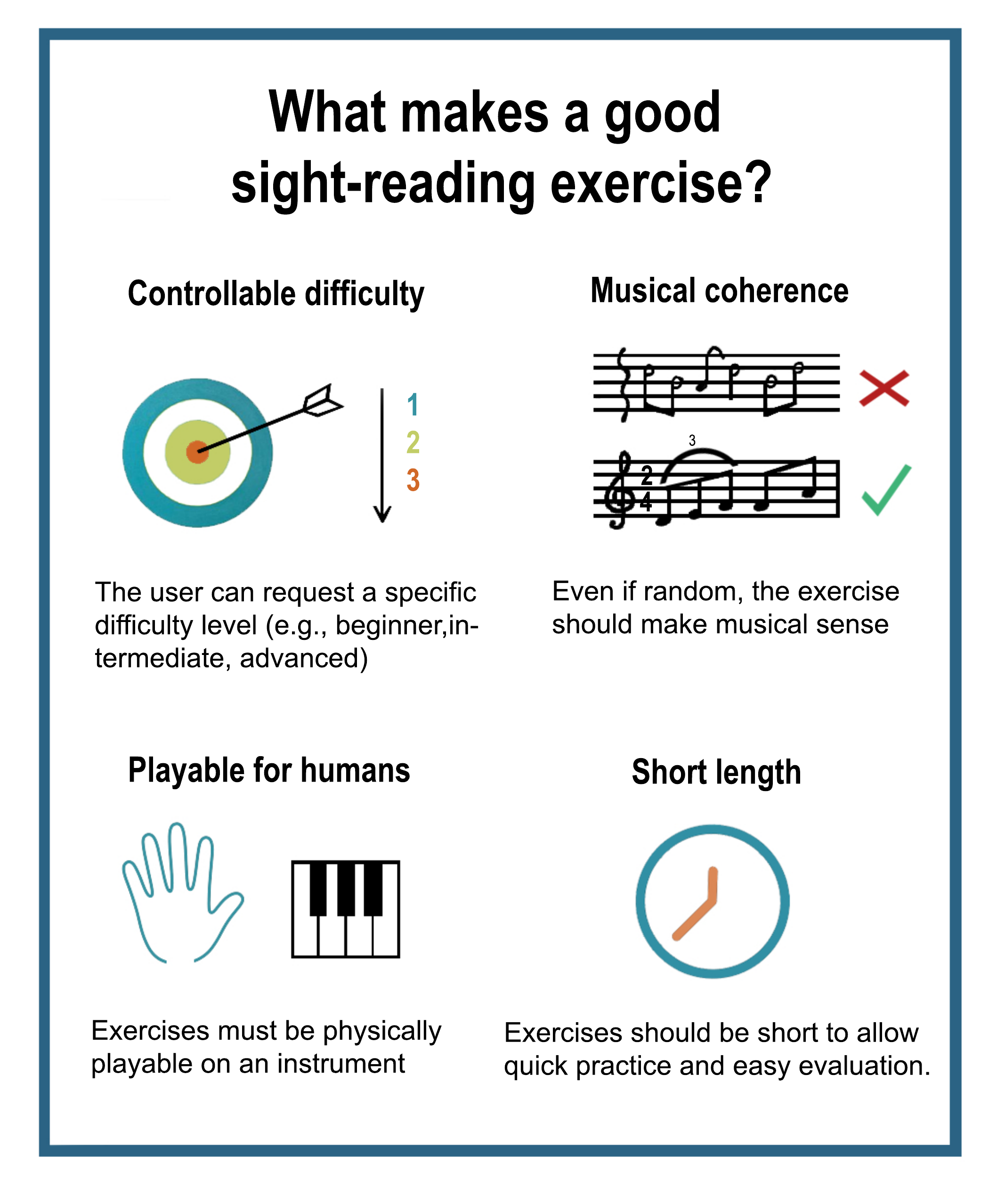}
    \caption{Effective sight-reading exercises must be brief, coherent, playable, and appropriate for the player's level. Our system explicitly controls difficulty while implicitly addressing the other constraints.}
    \label{fig:teaser}
\end{figure}

We introduce a system designed to produce \textit{personalized sight-reading exercises}. Personalization involves crafting suitable music compositions that are tailored to each piano student. To that end, and as a first step of control, we take advantage of recent advances in piano difficulty prediction~\cite{ramoneda2024combining,ramoneda2024difficulty} to enable conditioning on different difficulty levels. In addition to difficulty control, effective sight-reading exercises---as illustrated in Figure~\ref{fig:teaser}---must exhibit musical coherence: they should be readable by humans, playable on the instrument, and concise enough to fit typical practice routines. By combining autoregressive generation, difficulty conditioning, and multi-target optimization, we produce coherent and level-appropriate scores that adapt to the evolving needs of each learner. This approach enables structured, scalable generation aligned with the pedagogical goals in music education. Although previous efforts have addressed AI-assisted evaluation of sight-reading exercises~\cite{cheng2008nmf,huang2019sightreading}, the generation of pedagogically appropriate sight-reading material remains largely unexplored.

\begin{figure}[t]
    \centering
    \includegraphics[width=0.6\linewidth, trim=10pt 20pt 15pt 15pt, clip]{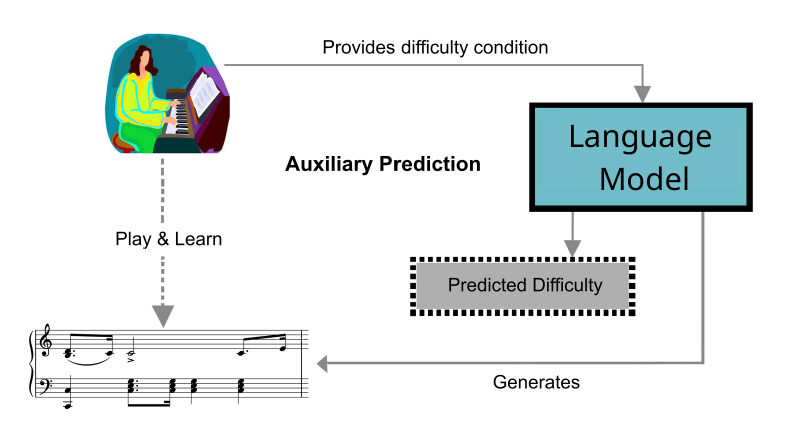}
    \caption{Overview of the proposed system for generating piano exercises with controllable difficulty. The model is trained autoregressively on symbolic music and conditioned on target difficulty levels. An auxiliary difficulty prediction task is included during training to encourage representations aligned with difficulty, although it is not used at generation time.}
    \label{fig:methods}
\end{figure}

Recent advances in symbolic music generation with language models have shown strong capabilities in producing coherent and stylistically controlled compositions~\cite{huang2019musictransformer,huang2020popmusictransformer,ma2021ailyricist}. However, most previous work has focused on MIDI generation~\cite{zhang2022structure,wang2022songdriver,ma2022userpreference,hu2022repetition,zhang2024agedm}, which is not suitable for educational applications because MIDI files are made to produce sound, not to be read by musicians. 
Unlike MusicXML, they cannot provide appropriate scores for educational practice. To generate exercises that are immediately usable by students and teachers, we directly produce MusicXML scores~\cite{good2021musicxml}, a widely adopted open standard supported by educators, notation software, and digital archives. Using a previously introduced tokenization method~\cite{suzuki2021scoretransformer}, we finetune pretrained large language models on MusicXML data to enable high-quality, legible score generation.

The generation of personalized sight-reading material can be framed as a sequential decision-making task, where the model must optimize for musical coherence, stylistic consistency, playability, and alignment with a target difficulty level~\cite{ramoneda2024combining,ramoneda2024difficulty}. Previous research has controlled autoregressive models by adding conditioning tokens at the beginning of the sequence~\cite{li-liang-2021-prefix}. In this work, we explore different strategies to encode the difficulty signal. However, complex conditioning variables, such as difficulty, do not always improve the next-token prediction. This issue, known as conditioning collapse, arises when control signals are partially or fully ignored during generation~\cite{liu2024controllable,lee2024fewshot,skopek2023towards,dathathri2019plug}. To mitigate this, we introduce an auxiliary difficulty prediction objective~\cite{li2020aux,yu2024aux}, illustrated in Figure~\ref{fig:methods}, which empirically improves the alignment between generated sequences and the target difficulty. 
This auxiliary loss also acts as a regularizer that reinforces the effect of the control signal. 

\paragraph{Our main contributions are:}
\begin{itemize}
    \item We present the first generative language model specifically designed to generate personalized sight-reading exercises in MusicXML format.
    \item We adapt pre-trained NLP models to symbolic music through fine-tuning and explore various prompt strategies---including zero-shot and chain-of-thought prompting---to align textual difficulty conditions with musical structure.
    \item We show that conditioning alone is insufficient to control global properties like difficulty, and propose an auxiliary difficulty prediction loss to regularize conditioning, to allow goal-directed generation aligned with pedagogical goals.
    \item We perform both objective evaluations and a user study to assess the effectiveness of our methodology, demonstrating its impact on the controllability, perceived difficulty, and educational relevance of the generated material.
\end{itemize}

This study investigates the use of adaptive sight-reading generation in tailoring music education through controllable models, thereby opening new directions in curriculum-aware and difficulty-conditioned music generation, highlighting the potential of general-purpose models for structured, human-centered content creation. We will release code and models to foster progress in this direction~\footnote{Code and models will be released upon paper acceptance.} and a companion page with examples~\footnote{At: \url{https://sight25.github.io}}.

\section{Literature Review}

This section reviews the research foundational to our personalized sight-reading system, focusing on score representation and difficulty prediction. 

\subsection{Score representations}
\label{ssec:lit_rep}

Symbolic music representations are a key component in most computational music systems. Early approaches focused on MIDI or piano roll formats due to their simplicity and availability \cite{huang2019musictransformer,huang2020popmusictransformer,hsiao2021compound}. However, these formats lack important information such as notational structure, visual layout, or expressive markings, essential for tasks such as music education, performance, or score editing.

To overcome the limitations of MIDI and piano roll, recent works adopt richer notation formats such as MusicXML, which preserve both musical semantics and visual aspects of the score \cite{musicxml40}. Other approaches use discrete token-based representations tailored for sequential models. For example, Score Transformer proposes a linearized token format for the generation of piano music \cite{suzuki2021scoretransformer}, while Measure by Measure introduces a hierarchical grid structure based on side bars \cite{yan2024measure}. ABC notation, a compact text-based format, offers a simpler alternative that supports fast composition modeling and is increasingly used in folk and choral music generation, as seen in models such as FolkRNN \cite{sturm2016music}, Tunesformer \cite{wu2023tunesformer} or DeepChoir \cite{wu2023chord}. In addition to symbolic processing tasks, transcription tasks such as Optical Music Recognition or audio-to-score conversion also benefit from richer notation formats, including Humdrum \cite{huron1995humdrum} and MusicXML variants \cite{yan2024measure}, which provide structured and expressive representations to align audio, visual, and symbolic data.

Among the different representations, we adopt the one proposed by Suzuki et al.~\cite{suzuki2024rearrangement}, as it is based on MusicXML and aligns with our goal of generating readable and editable sheet music. MusicXML is one of the most widely used formats in music education, notation software, and digital archives, making it a suitable choice for data-intensive machine learning tasks.
Our tokenization method preserves structural details such as measures, voices, and staves, as well as essential musical information including note names, durations, ties, stems, beams, and articulations. By converting these musical elements into linear token sequences, we facilitate efficient score modeling while retaining musical coherence.

\subsection{Difficulty prediction}
\label{ssec:lit_diff}

In this subsection, we review previous approaches to predicting musical difficulty, as obtaining a sufficiently large, expert-annotated dataset would be prohibitively expensive. Given that our methodology relies on automatically predicted synthetic labels, we survey existing methods for difficulty estimation to motivate our design choices and ensure methodological consistency throughout our approach.

Organizing educational material into steps of increasing difficulty, where mastering each step unlocks the next, is essential for effective learning. Playing an instrument involves not only reading the score but also body movement, sound production, and interpretative principles tied to specific performance traditions~\citep{cook1999analysing}. Consequently, performance difficulty can be understood as the effort required by a pianist to meet the expressive, technical, and notational demands of a passage~\citep{ramoneda2024combining}. Because objective criteria are lacking, difficulty assessments often involve subjectivity, leading to disagreements among educators and classification systems~\cite{wareham1967development}. Despite these variations, rankings from music boards, conservatories, composers, and publishers reveal general trends~\cite{ramoneda2024combining,zhang2023symbolic,ramoneda2024difficulty}.

\begin{table}[t!]
\small
\centering
\begin{tabular}{p{0.32\columnwidth}p{0.17\columnwidth}p{0.17\columnwidth}}
\cline{2-3}
                                & \multicolumn{2}{c}{CIPI Dataset}                  \\ \cline{2-3} 
                                & \multicolumn{1}{c}{Acc-9} & \multicolumn{1}{c}{MSE} \\ \hline
argnn variant~\cite{ramoneda2024combining}       & 32.6 (2.8)              & 2.1 (0.2)              \\
virtuoso model~\cite{ramoneda2024combining}      & 35.2 (7.3)              & 2.1 (0.2)              \\
pitch-feat model~\cite{ramoneda2024combining}    & 32.2 (5.9)              & 1.9 (0.2)              \\
deep ensemble~\cite{ramoneda2024combining}       & 39.5 (3.4)              & \textbf{1.1 (0.2)}     \\
RubricNet~\cite{ramoneda2024difficulty}         & \textbf{41.4 (3.1)}     & 1.7 (0.5)              \\
\hdashline
Gaussian Naive Bayes                & 38.7 (4.2)              & 1.8 (0.3)              \\
\hline
\end{tabular}
\caption{Comparison of classification accuracy (Acc-9) and mean squared error (MSE) on the CIPI dataset.}
\label{tab:cipi_results}
\end{table}

Early approaches relied on hand-crafted features extracted from symbolic scores, such as note density, pitch range, and articulation, combined with classical or rule-based machine learning models~\cite{sebastien2012score,chiu2012study}. Later studies incorporated performance-based features, such as the frequency of finger movement, to better model technical demands~\cite{nakamura2014merged,nakamura2015automatic}. More recent work explored deep learning models trained on annotated datasets~\cite{ramoneda2024combining,ramoneda2024difficulty}, which achieve strong predictive performance, but often lack interpretability.

To improve explainability, RubricNet~\cite{ramoneda2024difficulty} introduced a set of descriptors inspired by pedagogical rubrics, capturing aspects such as pitch range, entropy, rhythmic density, hand displacement, and structural repetitiveness. In this work, we adopt these descriptors and apply a lightweight \emph{Gaussian Naive Bayes} classifier to predict synthetic difficulty labels. This choice follows the RubricNet framework, but offers a simpler, more efficient, and reusable alternative for large-scale labeling. As shown in Table~\ref{tab:cipi_results}, this approach remains competitive while maintaining interpretability and calibration.

\subsection{Related generative work on difficulty control}

Several studies have also addressed difficulty-controlled score rearrangement. Nakamura and Yoshii~\cite{nakamura2018reduction} formulate piano reduction as a statistical optimization problem balancing difficulty and fidelity via fingering models. Gover and Zewi~\cite{gover2022amusic} adopt a MIDI translation approach to generate piano arrangements at different playing levels. Suzuki et al.~\cite{suzuki2024rearrangement} propose a notation-to-notation transformer model that modifies existing piano scores while preserving expressive marks. Unlike these approaches, we generate novel, sight-reading exercises from scratch, conditioned on interpretable difficulty descriptors. This design enables fine-grained control and better pedagogical alignment, bridging the gap between music informatics and music education.

\section{Methodology}

Our method for generating difficulty-aware symbolic music consists of three stages: (1) self-supervised pretraining using next-token prediction, (2) difficulty conditioning with various prompt strategies, and (3) refinement of the conditioning with an auxiliary loss.

\begin{figure}[t]
  \centering
  \includegraphics[width=0.4\linewidth, trim=10pt 20pt 15pt 50pt, clip]{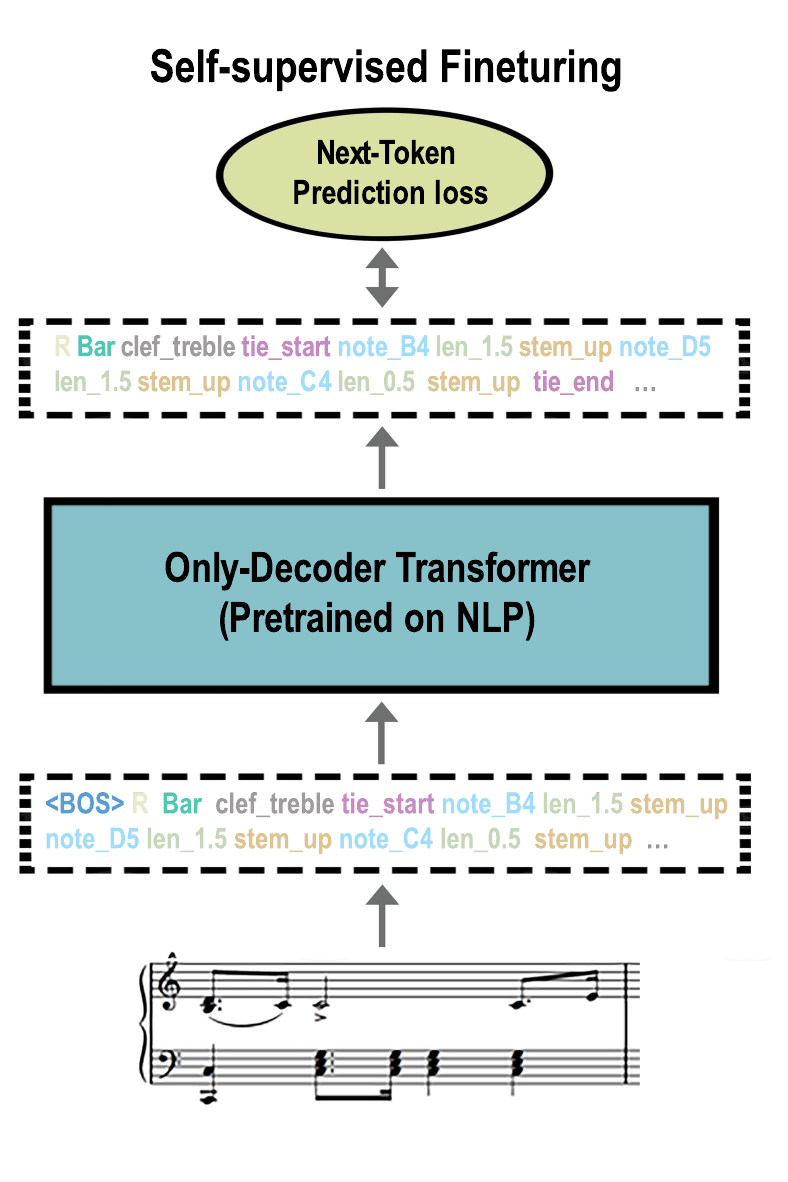}
  \caption{Self-supervised pretraining of a decoder-only Transformer on a large collection of music scores using next-token prediction. This stage allows the model to learn general musical structure.}
  \label{fig:step1}
  
\end{figure}

\subsection{Self-supervised pretraining}
\label{ssec:pretraining}

We pretrain a decoder-only Transformer using an autoregressive objective on symbolic music tokens, as shown in Figure~\ref{fig:step1}. The model uses causal masking (masked self-attention) to prevent access to future tokens during training. The representation we adopt is based on Suzuki et al.~\cite{suzuki2024rearrangement}, as introduced in Section~\ref{ssec:lit_rep}. Each score is linearized into a sequence of discrete tokens (e.g., note names, durations, articulations, markers), enabling the use of standard language modeling techniques.

Let \( x = (x_1, x_2, \dots, x_T) \) be a sequence of tokens. The model estimates the probability of the sequence via causal factorization:
\begin{equation}
    P(x) = \prod_{t=1}^{T} P(x_t \mid x_{<t}; \theta_{lm}),
\end{equation}
where \( \theta_{lm} \) are the model parameters.

We minimize the negative log-likelihood of the correct token at each timestep, resulting in the standard cross-entropy loss:
\begin{equation}
    \mathcal{L}_{\text{CE}} = -\sum_{t=1}^{T} \log P(x_t \mid x_{<t}; \theta_{lm}).
\end{equation}

We experiment with two families of architectures pre-trained in text: LLaMA3 models \cite{llama3herd} and the HuggingFace Smol v2 \cite{benallal2024smollm2} series. To enable efficient fine-tuning, we use low-rank adaptation (LoRA) \cite{hu2022lora}, a parameter-efficient technique that injects trainable low-rank matrices into the frozen weights of the pre-trained model.

Formally, for a weight matrix \( W \in \mathbb{R}^{d \times k} \), LoRA introduces two trainable matrices \( A \in \mathbb{R}^{d \times r} \) and \( B \in \mathbb{R}^{r \times k} \), with \( r \ll \min(d, k) \). During finetuning, the weight is updated as:
\begin{equation}
    W' = W + \alpha AB,
\end{equation}
where \( \alpha \) is a scaling factor. This reduces the number of trainable parameters while preserving performance.

\begin{figure}[t]
  \centering
  \includegraphics[width=0.45\linewidth, clip]{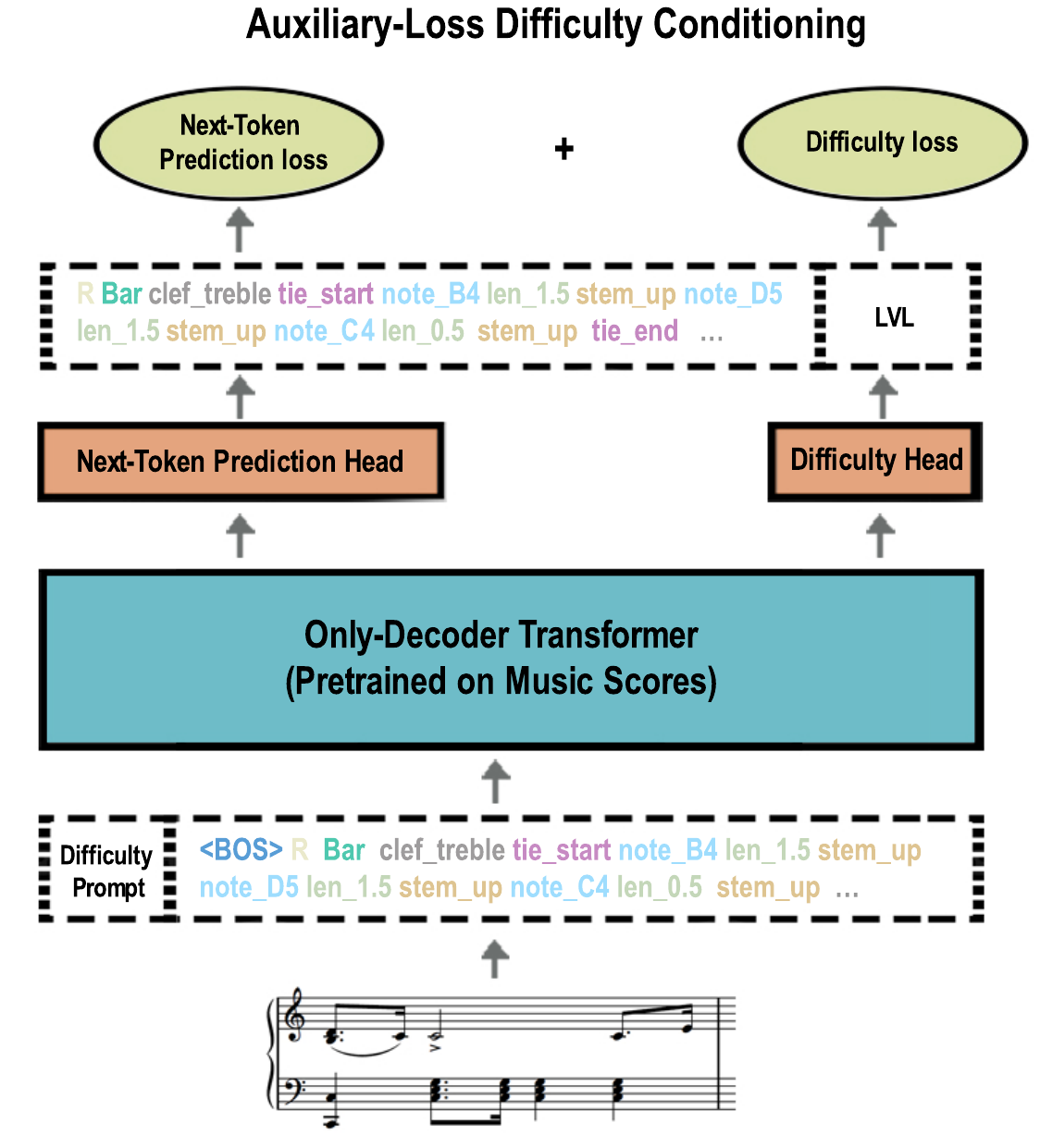}
  \caption{Difficulty conditioning through an auxiliary loss that encourages the model to align generation with target difficulty levels, enabling controllable music generation.}

  \label{fig:step2}
\end{figure}

\subsection{Difficulty conditioning}
\label{ssec:conditioning}

\begin{figure*}[ht!]
\scriptsize
\centering
\renewcommand{\arraystretch}{1.2}
\begin{tabular}{p{0.15\textwidth} p{0.75\textwidth}}
\toprule
\textbf{Type} & \textbf{Prompt Content} \\
\midrule
\texttt{diff} & \texttt{Easy} \\
\texttt{diff\_cot} & \texttt{To compose a piano piece, we start by defining its characteristics. An Easy piece should have a clear melody, limited movement, and a steady rhythm. A Mid piece introduces a wider pitch range, moderate hand movement, and rhythmic variety. An Advanced piece demands technical skill, with complex passages, fast sequences, and wide hand displacement. Now, generate a Easy piano composition in MusicXML format, ensuring structure and musical coherence.} \\
\texttt{feats} & \texttt{Easy: 0.21 0.19 0.35 0.28 0.50 0.45 0.60 0.58 0.33 0.31 0.27 0.29} \\
\texttt{feats\_cot} & \texttt{We will generate a Easy piano piece. First, we analyze its musical descriptors. Pitch entropy shows pitch variety. RH: 0.21, LH: 0.19. Higher = more diverse notes. Pitch range is the distance between lowest and highest notes. RH: 0.35, LH: 0.28. Avg. pitch gives the central register. RH: 0.50, LH: 0.45. Higher = brighter. Displacement rate reflects hand movement intensity. RH: 0.60, LH: 0.58. Avg. IOI is the time between note onsets. RH: 0.33, LH: 0.31. Lower = denser rhythm. Pitch set LZ shows structural complexity and repetitiveness. RH: 0.27, LH: 0.29. Now generate a coherent and structured piece in MusicXML.} \\
\bottomrule
\end{tabular}
\caption{Examples of the four prompt types used to encode difficulty. Each strategy varies in length, explicitness, and semantic richness, from simple zero-shot labels to detailed chain-of-thought rationales.}
\label{fig:prompt-examples}
\end{figure*}

To condition the model on difficulty, we adopt a prepending strategy, where a natural language prompt indicating the target difficulty is placed at the beginning of each sequence, as shown in Figure~\ref{fig:step2}. This allows the model to adjust its generation according to the desired difficulty level while preserving coherence and musical structure.

The models used in this phase are those pre-trained in the self-supervised stage and finetuned using LoRA, as described in Section~\ref{ssec:pretraining}. This approach enables efficient adaptation to the MusicXML structure while retaining the language knowledge of the original model. Thanks to this, we can use natural language prompts (e.g., \texttt{"Easy"}) to guide the generation process.

To minimize context length, self-supervised pretraining was performed with an extended vocabulary that included MusicXML-specific symbols, incorporating the tokenization vocabulary introduced in Section~\ref{ssec:lit_rep}. However, we did not modify the original natural language vocabulary. During the difficulty conditioning stage, we finetune both vocabularies jointly: the music vocabulary and the natural language vocabulary.

Given the sequence \( x \) and a prompt \( p \), the conditioning is modeled as:
\begin{equation}
    P(x \mid p) = \prod_{t=1}^{T} P(x_t \mid p, x_{<t}; \theta_{lm}), 
\end{equation}
where \( \theta_{lm} \) are the parameters of the only-decoder transformer and the loss remains the standard autoregressive cross-entropy loss. In the computation of \( \mathcal{L}_{lm} \), we ignore the prompt \( p \).

For each training fragment \( x \), we compute a difficulty label \( d(x) \in \mathbb{N}_0 \) and optionally features \( f(x) \in \mathbb{R}^n \), as described in Section~\ref{ssec:lit_diff}. 
These labels are \emph{synthetic} and obtained using the Gaussian Naive Bayes model introduced in Section~\ref{ssec:lit_diff}. 
They are then used to build the prompt \( p(x) {\scriptstyle =} g(f(x), d(x), t) \), where \( t \in \{\texttt{diff},\allowbreak\ \texttt{diff\_cot},\allowbreak\ \texttt{feat},\allowbreak\ \texttt{feat\_cot}\} \) denotes the prompt type used for fine-tuning.

The \texttt{diff} format is a minimal, zero-shot prompt containing only the difficulty label \( d(x) \) (e.g., \texttt{Advanced}). The \texttt{diff\_cot} format extends this with a natural-language description of what difficulty implies, following a chain-of-thought strategy to enhance reasoning~\cite{wei2022chain}. The \texttt{feat} format encodes only the normalized extracted features \( f(x) \), while \texttt{feat\_cot} combines \( f(x) \) with a verbalized explanation, also following a chain-of-thought strategy. Figure~\ref{fig:prompt-examples} provides examples of these prompt formats.

\subsection{Multi-objective optimization with an auxiliary loss}
\label{ssec:aux}

Difficulty is a complex and subjective concept and does not align well with the cross-entropy loss described in the previous section. In practice, generating with difficulty conditioning using only the main autoregressive objective can cause \emph{conditioning collapse}, where the model learns to ignore the conditioning signal, either fully or partially. This issue has been observed in other generation tasks that involve style, sentiment, or structure. To address this, previous work has explored solutions such as auxiliary objectives~\cite{zhu2023fine,li2020aux,yu2024aux,wo2025aux}, latent space interpolation~\cite{lee2024fewshot}, constraint-based refinement~\cite{skopek2023towards}, and inference-time control~\cite{dathathri2019plug}. These studies show that autoregressive loss alone is not enough to ensure that the model respects complex conditioning, especially when dense supervision is lacking. Following this line, we introduce an auxiliary task that explicitly models the difficulty. The effectiveness of using auxiliary objectives to improve conditioning has been clearly demonstrated by prior studies in other applications~\cite{li2020aux,yu2024aux,wo2025aux},
and in our case, this additional objective helps the model better match the intended difficulty level.

Initially, the input sequence \( x \) is preceded by a natural language prompt \( p \), which encodes the desired difficulty level, as outlined in Section~\ref{ssec:conditioning}. Second, we add an auxiliary prediction head at the final token position (denoted \texttt{END}) that classifies the target difficulty \( d \). This head operates on the final hidden state of the sequence and is trained to predict the intended difficulty class.

Let \( d \) be the target difficulty label (e.g., easy, medium, advanced). We define the auxiliary classification loss as:
\begin{equation}
    \mathcal{L}_{\text{diff}} = -\log P(d \mid x; \theta_{\text{diff}}),
\end{equation}
where \( \theta_{\text{diff}} \) are the parameters of the difficulty classification head.

The total training loss is a weighted combination of the autoregressive and auxiliary objectives:
\begin{equation}
    \mathcal{L}_{\text{total}} = \mathcal{L}_{\text{CE}} + \beta \cdot \mathcal{L}_{\text{diff}},
\end{equation}
where \( \beta \) is a scalar that controls the influence of the auxiliary loss.

To reduce the risk of shortcuts---such as the model copying the difficulty label from the prompt into the final classification head---we detach the gradient flow from the auxiliary classification loss \( \mathcal{L}_{\text{diff}} \) into the parameters \( \theta_{\text{LM}} \) of the language model. In other words, during backpropagation, only the classification head parameters \( \theta_{\text{diff}} \) are updated by \( \mathcal{L}_{\text{diff}} \), while \( \theta_{\text{LM}} \) remains unaffected by this loss: \( \nabla_{\theta_{\text{LM}}} \mathcal{L}_{\text{diff}} = 0 \), and \( \nabla_{\theta_{\text{diff}}} \mathcal{L}_{\text{diff}} \neq 0 \). This design choice aims to discourage trivial solutions based on surface patterns in the prompt and instead promote the encoding of difficulty information within the generated content. Although the auxiliary head does not directly influence the generation, it provides a weak supervision signal that may help to reinforce the consistency between the output and the intended difficulty level.

\section{Experimental Setup}

\subsection{Data}

We use the PDMX dataset, recently introduced by Long et al.~\cite{long2024pdmx}, which contains more than 200,000 public domain MusicXML scores collected from the MuseScore platform. PDMX is the largest copyright-free MusicXML dataset available and includes rich metadata such as user tags and ratings and we use it as our main training corpus.~\footnote{We train directly on MusicXML, the most widely used open format among users, since alternative formats like MIDI and ABC either lack essential notation details (e.g., pitch spelling, beam grouping, voice separation) and have limited reliable conversion tools to MusicXML.} From this collection, we extract 41,170 piano pieces with exactly two staves and high community ratings. We segment each piece into 16-bar fragments and estimate their difficulty using the method of Section~\ref{ssec:lit_diff}, which predicts both interpretable descriptors and difficulty levels. We group the data into three difficulty classes: Easy (level 0), Medium (level 1), and Advanced (level 2 or higher). Although the PDMX dataset is skewed toward easier fragments (61.5\% Easy, 37.0\% Medium, 12.1\% Advanced), we balance the samples between difficulty classes during fine-tuning to avoid bias in the generative model. Similarly, for evaluation, we generate an equal number of samples per class to ensure a fair comparison.

\begin{table}[t]
\centering
\small
\begin{tabular}{lcc}
\toprule
\textbf{Dataset} & \textbf{Pieces} & \textbf{Tokens (raw)} \\
\midrule
\textbf{PDMX Train} & 33,962 & 45,764,701 \\
\textbf{PDMX Validation} & 7,176 & 10,026,819 \\
\textbf{PDMX Train (augmented)} & 505,838 & 659,061,903 \\
\textbf{Polish Classical} & 1,532 & 4,581,980 \\
\textbf{Craig Sapp Classical} & 852 & 5,406,551 \\
\textbf{Internal (J-Pop)} & 1,061 & 6,192,391 \\
\bottomrule
\end{tabular}
\caption{Model performance (lower is better) across next token prediction pretraning. \textit{NLP Pretrain} indicates whether the model was pretrained on language data. \textit{Steps} shows the number of training steps needed for convergence.}
\label{tab:dataset_stats}
% \vspace{-0.75cm}
\end{table}

% \begin{table*}[ht!]
%     \centering
%     \tiny
%     \begin{tabular}{lccc|cccc}
%         \toprule
%         \textbf{Model} & \textbf{Params} & \textbf{NLP Pretrain} & \textbf{Steps} & \textbf{PDMX Val} & \textbf{Polish Classical} & \textbf{Craig S. Classical} & \textbf{Internal J-Pop}  \\
%         \midrule
%         GPT\-2 & 100M & \xmark & 120k & 0.38 & 0.64 & 0.66 & 1.21 \\
%         \midrule
%         LLaMA (3.2) & 1B & \cmark & 50k & 0.37 & 0.62 & 0.63 & 0.97 \\
%         LLaMA (3.1) & 8B & \cmark & 40k & \textbf{0.36} & \textbf{0.62} & \textbf{0.60} & \textbf{0.83} \\
%         \midrule
%         Smol & 135M & \cmark & 70k & 0.38 & 0.63 & 0.64 & 1.04 \\
%         Smol & 1.7B & \cmark & 30k & 0.38 & 0.62 & 0.63 & 1.07 \\
%         \bottomrule
%     \end{tabular}
%     \caption{Model performance (lower is better) across difficulty prediction tasks. \textit{NLP Pretrain} indicates whether the model was pretrained on language data. \textit{Steps} shows the number of training steps needed for convergence.}
%     \label{tab:results}
% \end{table*}

We divide the 41,170 scores into 80\% for training and 20\% for validation. To increase diversity and improve robustness to key changes, we apply automatic transpositions of $-6$ and $+6$ semitones to the training set using MuseScore~\cite{musescore45nightly}, adjusting the key signature accordingly. To avoid introducing noisy or rare notation, we discard any transposed samples that result in double sharps or double flats. This results in a clean and musically consistent training set enriched with a wide range of tonal contexts. The final augmented training set contains 505,838 unique piano pieces and more than 659 million symbolic tokens. We used this expanded data set during self-supervised pretraining to improve generalization and robustness.

To evaluate generalization beyond the biased distribution of PDMX, we include three additional test sets in the self-supervised pretraining, as shown in Table~\ref{tab:dataset_stats}. These consist of two public datasets and one private collection: (1) a set of Polish classical scores, (2) the Craig Sapp Classical collection, and (3) a private dataset of Japanese pop music. These collections allow us to assess whether our models trained in PDMX can generalize to diverse styles and distributions.

\subsection{Training details}

We tokenize all symbolic music using the vocabulary introduced in Section~\ref{ssec:lit_rep}, based on a MusicXML structure~\cite{suzuki2021scoretransformer}. To reduce noise and sparsity, we discard tokens that appear less than 50 times in the training set, resulting in a compact vocabulary that still captures the musical variety of PDMX. All experiments use LoRA adapters with rank 64 and alpha 128 to enable parameter-efficient finetuning with reduced memory usage. Initial tests with other common configurations showed similar results. Experiments are conducted on a single NVIDIA A100 GPU, using gradient accumulation when needed to fit large batches in memory.

We use the AdamW optimizer in all training steps and train until convergence. Early stopping is applied every two epochs, based on validation performance. For self-supervised pretraining, we monitor the validation loss on the validation split of PDMX. For conditioning strategies, we evaluate difficulty prediction accuracy on 1{,}000 generated samples. During pretraining, we use a cosine learning rate scheduler with a minimum value of 0.001 and a batch size of 128. For difficulty conditioning, we used a fixed learning rate of 0.0001 and a batch size of 16, as larger sizes did not show clear benefit. We set $\beta = 0.1$ for the auxiliary prediction loss.

\section{Results}

\subsection{Self-supervised finetuning}

\begin{table*}[t!]
    \centering
    \footnotesize
    \begin{tabular}{lccc|cccc}
        \toprule
        \textbf{Model} & \textbf{Params} & \textbf{Pret.} & \textbf{Steps} & \textbf{PDMX Val} & \textbf{Polish} & \textbf{Classical} & \textbf{J-Pop}  \\
        \midrule
        GPT-2 & 100M & \xmark & 120k & 0.38 & 0.64 & 0.66 & 1.21 \\
        \midrule
        LLaMA3 & 1B & \cmark & 50k & 0.37 & 0.62 & 0.63 & 0.97 \\
        LLaMA3 & 8B & \cmark & 40k & \textbf{0.36} & \textbf{0.62} & \textbf{0.60} & \textbf{0.83} \\
        \midrule
        Smol & 135M & \cmark & 70k & 0.38 & 0.63 & 0.64 & 1.04 \\
        Smol & 1.7B & \cmark & 30k & 0.38 & 0.62 & 0.63 & 1.07 \\
        \bottomrule
    \end{tabular}
\caption{Model performance (lower is better) during the self-supervised pretraining stage. 
\textit{NLP Pretrain} indicates whether the model was pretrained on language data. 
\textit{Steps} shows the number of training steps needed for convergence.}
    \label{tab:results}
\end{table*}

Table~\ref{tab:results} reports the validation loss in four benchmark datasets after the self-supervised pretraining stage in PDMX Train. All models were trained in the same symbolic music corpus using the cross-entropy objective described in Section~\ref{ssec:pretraining}. We include models of different sizes and pretraining strategies to compare generalization, convergence speed, and adaptation to symbolic music. In particular, we selected two families of pre-trained models (LLaMA3~\cite{llama3herd} and Smol~\cite{benallal2024smollm2}) to explore how language pretraining impacts symbolic music modeling, and we included GPT-2 as a baseline widely used in prior midi generation models~\cite{pasquier2025,thickstun2024anticipatory}.

The best overall performance is achieved by LLaMA 8B, which obtains the lowest validation loss across all datasets with only 40k training steps. This suggests that large-scale models pre-trained on text can efficiently adapt to symbolic music with relatively few updates. In contrast, GPT-2, trained from scratch without prior NLP pretraining, requires 120k steps and yields the highest validation loss, confirming the benefit of using pre-trained language models. Among smaller models, Smol 135M generalizes better than GPT-2 and converges faster (70k vs. 120k), while Smol 1.7B converges in just 30k steps but underperforms compared to the LLaMA variants. LLaMA 1B, trained for 50k steps, also exceeds all smaller models.

Interestingly, larger models show better generalization in the \textit{internal J-Pop} dataset, which is more out of distribution compared to the classical training data, suggesting higher robustness to domain shift. Despite this, Smol 135M achieves competitive results on all benchmarks. Moreover, it is based on an efficient architecture, which enables efficient deployment on CPUs and consumer GPUs. In particular, inference on a laptop GPU (e.g., 3050) takes $<$30s per piece, which is suitable for sight-reading educational use case. Therefore, we opted to employ Smol 135M as our minimalistic benchmark model and LLaMA 1B as our superior performance reference model in our comparative study. We refrained from using a larger model in order to maintain efficiency and balance resource consumption with performance outcomes.

\subsection{Conditioning}

\begin{table}[t!]
\footnotesize
\centering
\resizebox{.45\columnwidth}{!}{%
\begin{tabular}{llccc|cc}
\toprule[1pt]
\multicolumn{2}{l}{} & \multicolumn{3}{c|}{\textbf{Cond. Gen.}} & \multicolumn{2}{c}{\textbf{Diff.}} \\
\cmidrule(lr){3-5} \cmidrule(lr){6-6}
\textbf{   } & \textbf{Setup} & \textbf{Acc (\%)} & \textbf{MSE} & \textbf{Loss} & \textbf{Acc (\%)}\\
\midrule
\multicolumn{7}{l}{\textit{Smol (135M) baseline}} \\
& diff         & 69.62  & 0.38 & 1.32 & --    \\
& diff\_cot    & 69.33  & 0.30 & 1.34 & --    \\
& feats        & deg.   & deg. & 1.28 & --    \\
& feats\_cot   & 64.34  & 0.60 & 1.37 & --    \\
\midrule
\multicolumn{7}{l}{\textit{Smol (135M) Auxiliary loss}} \\
& diff         & 83.30  & 0.22 & 1.30 & 97.24  \\
& diff\_cot    & \textbf{92.88}  & \textbf{0.09} & 1.47 & 96.98  \\
& feats        & deg.   & deg. & 1.05 & 96.42 \\
& feats\_cot   & 67.17  & 0.54 & 1.39 & 97.25  \\
\midrule
\multicolumn{7}{l}{\textit{LLaMA 3 (1B) Auxiliary loss}} \\
& diff         & 87.21  & 0.24 & 1.01 & 91.22  \\
& diff\_cot    & 92.01    & 0.16   & 1.10   & 0.92  \\
& feats        & 0.51   & 0.92 & 1.02 & 54.88  \\
& feats\_cot   & 0.64   & 0.60 & 0.97 & 90.35  \\
\bottomrule[1pt]
\end{tabular}
}
\caption{Results of the conditioning experiments. The first three columns, \textit{Cond. Gen.} show the quality of conditional generation. The last one, \textit{Pred.},  report the performance of the auxiliary difficulty prediction task. \textit{deg.} indicates that the task degenerated.}
\label{tab:representation}
\end{table}

We evaluated the effectiveness of difficulty conditioning strategies and the auxiliary classification objective described in Sections~\ref{ssec:conditioning} and~\ref{ssec:aux}. Table~\ref{tab:representation} presents a comprehensive comparison in multiple setups. These include a 135M parameter model (\textit{Smol}) trained with and without the auxiliary loss, and a larger LLaMA 3 model of 1B parameters, always using the four prompting strategies introduced earlier: \texttt{diff}, \texttt{diff\_cot}, \texttt{feats}, and \texttt{feats\_cot} (see Figure~\ref{fig:prompt-examples}).

Each setup is evaluated along two axes: the quality of difficulty-conditioned generation and the accuracy of the auxiliary difficulty classification task. For the former, we compute the classification accuracy of the difficulty level and the mean squared error (MSE) of 1000 generated conditioned sequences, and the standard autoregressive cross-entropy loss in PDMX Val. For the latter, we report the prediction accuracy of the auxiliary classification head in PDMX Val. In cases where the model degenerated or the metric was not applicable, we report this explicitly.

The results show that using only the autoregressive loss (top block, Table~\ref{tab:representation}) leads to relatively poor conditioning. For example, when using the \texttt{diff\_cot} strategy without auxiliary loss, it reaches only 69.33\% precision, while the \texttt{feats} strategy even degenerates. This supports the hypothesis of partial conditioning collapse, indicating that cross-entropy loss alone is suboptimal for difficulty-based control. The signal is too weak to effectively help minimize the loss of the next token prediction.

In contrast, when the auxiliary difficulty classification loss is added (middle block, Table~\ref{tab:representation}), we observe large improvements across most metrics. With the \texttt{diff} prompt, generation accuracy improves from 69.62\% to 83.30\%, and with the more descriptive \texttt{diff\_cot} prompt, it reaches 92.88\%--the highest among all configurations. This supports the idea that chain-of-thought prompting helps models interpret difficulty intent more effectively. It also shows that the auxiliary head helps, improving both generation accuracy and feature alignment (MSE: 0.09). Furthermore, the classification accuracy of the auxiliary head is consistently above 96\%, indicating that the model recognizes the difficulty of the generated sequence.

In the LLaMA 3 model (bottom block, Table~\ref{tab:representation}), we observe strong generation performance even when using feature-based prompts (\texttt{feats\_cot}), showing that scaling up the model helps overcome the limitations of less explicit conditioning. The model reaches a generation accuracy of 64\% and a difficulty classification accuracy of 90.35\%, confirming that larger models are more robust to weaker signals. The \texttt{diff\_cot} experiment confirms our hypothesis that the chain-of-thought strategy improves performance. However, the results are similar to those of the Smol version, although with a better validation loss. The \texttt{feats} experiment did not degenerate, but shows very unstable behavior and poor results, highlighting the importance and effectiveness of the chain-of-thought strategy.

% \begin{table}[t!]
% \centering
% \resizebox{.85\columnwidth}{!}{%
% \begin{tabular}{llccc|cc}
% \toprule[1pt]
% \multicolumn{2}{l}{} & \multicolumn{3}{c|}{\textbf{Cond. Gen.}} & \multicolumn{2}{c}{\textbf{Diff.}} \\
% \cmidrule(lr){3-5} \cmidrule(lr){6-6}
% \textbf{   } & \textbf{Setup} & \textbf{Acc (\%)} & \textbf{MSE} & \textbf{Loss} & \textbf{Acc (\%)}\\
% \midrule
% \multicolumn{7}{l}{\textit{Smol (135M) baseline}} \\
% & diff         & 69.62  & 0.38 & 1.32 & --    \\
% & diff\_cot    & 69.33  & 0.30 & 1.34 & --    \\
% & feats        & deg.   & deg. & 1.28 & --    \\
% & feats\_cot   & 64.34  & 0.60 & 1.37 & --    \\
% \midrule
% \multicolumn{7}{l}{\textit{Smol (135M) Auxiliary loss}} \\
% & diff         & 83.30  & 0.22 & 1.30 & 97.24  \\
% & diff\_cot    & \textbf{92.88}  & \textbf{0.09} & 1.47 & 96.98  \\
% & feats        & deg.   & deg. & 1.05 & 96.42 \\
% & feats\_cot   & 67.17  & 0.54 & 1.39 & 97.25  \\
% \midrule
% \multicolumn{7}{l}{\textit{LLaMA 3 (1B) Auxiliary loss}} \\
% & diff         & 87.21  & 0.24 & 1.01 & 91.22  \\
% & diff\_cot    & 92.01    & 0.16   & 1.10   & 0.92  \\
% & feats        & 0.51   & 0.92 & 1.02 & 54.88  \\
% & feats\_cot   & 0.64   & 0.60 & 0.97 & 90.35  \\
% \bottomrule[1pt]
% \end{tabular}
% }
% \caption{Results of the conditioning experiments. The first three columns, \textit{Cond. Gen.} show the quality of conditional generation. The last one, \textit{Pred.},  report the performance of the auxiliary difficulty prediction task. \textit{deg.} indicates that the task degenerated.}
% \label{tab:representation}
% \end{table}

We observe that larger models achieve consistently lower generation loss but perform worse in the auxiliary difficulty prediction task. This pattern appears in several setups, including \texttt{diff}, \texttt{diff\_cot}, and \texttt{feats\_cot}. For example, in \texttt{diff}, the LLaMA 3 model reaches a generation loss of 1.01 but only 91.22\% accuracy in difficulty prediction, while the smaller model achieves higher accuracy (97.24\%) despite a higher loss (1.30). A similar trend is observed in \texttt{diff\_cot} and \texttt{feats\_cot}. Including the auxiliary head improves controllability and reduces conditioning collapse, even though its gradients are detached from the language model parameters. Previous studies in conditional text generation \cite{li2020aux,yu2024aux,wo2025aux} have shown that auxiliary tasks can indirectly shape internal representations, since the model must maintain hidden states that are simultaneously predictive of the next token and separable across difficulty classes. We hypothesize that, in our case, this auxiliary head acts as a regularizer that biases shared representations toward encoding difficulty-relevant features, thereby reinforcing the effect of conditioning. Future work will investigate this alignment more directly through representation analyses.

We also conducted experiments replacing the natural-language descriptions in \texttt{diff\_cot} prompts with randomized text of the same length. Using the smaller \textit{Smol-v2} model, this caused the task to degenerate, showing that semantic content is crucial. Moreover, although \texttt{feats\_cot} prompts are longer than \texttt{diff\_cot}, the latter performs better, suggesting that the gains come from linguistic priors and semantic alignment rather than prompt length. Finally, we tested the auxiliary objective without gradient detachment. In this case, the classification head quickly overfit by exploiting shortcuts from the prompt tokens, and controllability did not improve.

In general, these results confirm the benefits of combining explicit prompting strategies with a weak supervision signal from an auxiliary task. In particular, chain-of-thought formulations and the auxiliary classification head reinforce each other, improving the alignment between generated music and the intended difficulty level. In particular, larger models such as LLaMA 3 show strong performance even with the same scarce training data, suggesting that scaling up the model can compensate for limited supervision and weaker prompt structures. However, more research is needed to understand how to incorporate the composite signals in \texttt{feats} and \texttt{feats\_cot} more effectively into the conditioning process.

% EXPLICAR LOS EXPERIMENTOS QUE HEMOS LLEVADO A CABO y explicar las metricas

% ANALIZAR COMO el COT suele ir mejor que no usar el COT en mse y acc DECIR QUE en FEAT DIRECTAMENTE SIN COT NO FUNCIONA 

% COMO EL MODELO DE 1B pese a tener peor difficulty validation saca unos resultados similares pero con ua validation loss mucho menor en mse y acc. LANZAR ALGUNA HIPOTESIS DE POR QUE ESTA PASANDO ESTO  Y QUE SIGNIFICA. HABLAR UN POCO DE QUE GARANTIZAR DE QUE NO HAY NINGUN SHORCUTTINEING QUEDA FUERA DE NUESTRO RESERCH PERO QUE PODRIA EXPLICAR POR QUE SACA ESO ASI

% HABLAR SOBRE LAS  DEGENERACIONES

% Some of the experiments resulted in degenerate outputs (e.g., empty sequences, format errors, or lack of interpretable difficulty cues), which made it impossible to compute evaluation metrics such as accuracy or mean squared error. In Table~\ref{tab:representation}, these cases are marked as \textit{deg.}, indicating that the output was degenerate and excluded from the calculation of the metric. DECIR QUE EN EL LLAMA 1b NO HAY DEGENERACIONES PERO COMO SE PUEDE VER FEAT ES MUY INESTABle

% \subsection{Conditioning in features}

\section{User Study}

We conducted a user study with five pianists.
The participants have piano training and 25-40 years of experience (average 35 years). Each pianist self-reported strong sight-reading skills, indicating that they can learn and perform a simple one-page piece (i.e., difficulty comparable to Burgmüller's \textit{25 Etudes}) with less than one hour of practice. They also expressed confidence in their ability to estimate the performance difficulty of a piece simply by examining the score. Expertise was prioritized over quantity, as assessing difficulty requires deep pedagogical and
performance knowledge. Recruiting such profiles is challenging and involving five experts already
adds strong value. 

\subsection{Experimental conditions}

We prepared twelve musical excerpts, each consisting of 16 measures. Specifically, six pieces were sampled from ground truth data, two pieces for each of the three target difficulty levels (denoted \textit{Oracle}).
Likewise, six pieces were generated using the LLaMA 3 Auxiliary Loss Model Diff (1B), with two pieces for each target difficulty level (denoted \textit{Proposed}).

Each participant was instructed to sight-read each piece.
Participants knew the purpose of the study, but did not know the number of compared methods or whether the score was created by humans.
The pieces were presented randomly, with neither the methodology nor the difficulty level being given.
Following each performance, they were asked to rate three aspects of the score on a five-point Likert scale:

\begin{enumerate}
\item \textbf{Readability}: The cognitive ease of sight-reading the score. Ratings range from 1 (requires substantial time to study the score) to 5 (very easy to read). Participants were instructed to account for notation quality as part of their assessment.
\item \textbf{Naturalness}: The degree to which the music feels musically coherent and idiomatic. Ratings range from 1 (musically incoherent) to 5 (entirely natural). The endings were intentionally truncated to sixteen bars, so the participants were asked to assess naturalness with this design choice in mind.
\item \textbf{Playability}: The technical ease of performing the piece after learning it, ranging from 1 (difficult) to 5 (easy).
\end{enumerate}

Participants were also asked to provide additional free-form comments on each piece.

\subsection{Results and discussions}

\subsubsection{Comparison of the questionnaires}

\begin{table}[t]
\centering
\begin{tabular}{lccc}
\toprule[1pt]
 & {\textbf{Readability}} & {\textbf{Naturalness}} & {\textbf{Playability}} \\
\hline
{\textit{Oracle}/}{Easy} & 4.60 & 3.90 & 4.20 \\
{\textit{Oracle}/}{Medium} & 2.80 & 3.80 & 3.10 \\
{\textit{Oracle}/}{Advanced}& 2.10 & 3.60 & 2.40 \\
{\textit{Proposed}/}{Easy} & 4.40 & 4.30 & 4.20 \\
{\textit{Proposed}/}{Medium} & 3.20 & 3.90 & 3.80 \\
{\textit{Proposed}/}{Advanced} & 3.60 & 3.90 & 3.50 \\
\bottomrule[1pt]
\end{tabular}
\caption{Ratings of the oracle and the proposed method across three difficulty levels.}
\label{tab:oracle_proposed_comparison}
\end{table}

Table~\ref{tab:oracle_proposed_comparison} presents the ratings for the oracle and the proposed methods across the three difficulty levels.
The playability ratings correspond inversely to the difficulty levels of the score. 
This indicates that scores conditioned to be more difficult are indeed perceived as more difficult to play -- even when the performer has learned the piece.  
Since the playability tends to be lower for the oracle as the difficulty increases, the proposed method creates easy scores for a given difficulty level.

Looking at the readability ratings, the proposed method seems to create pieces that are easier to sight-read than the oracle.  
The readability can be higher if the score is musically natural or the notation can be understood with little cognitive burden, but the consistently higher naturalness ratings suggest the former. The proposed method thus produces music that adheres more closely to familiar musical idioms than the oracle, making it easier to interpret and perform.  Additionally, the higher playability ratings of the proposed method compared to the oracle suggest that the proposed method often generates simpler compositions.

These results show that our method can generate pieces (1) whose difficulty controls are qualitatively noticeable, (2) make musical sense, and (3) are playable.  
Surprisingly, the proposed method generated scores that rated better than the oracle.  Although the underlying cause is unclear and requires further research, we observed that the oracle pieces tended to be more complex, as supported by the lower playability ratings, making proper rendering of MusicXML more difficult in music notation software.  This might contribute to the oracle scores being more technically demanding, while having notation quirks.

\subsubsection{Analysis of the comments}

\begin{figure}[t]
  \centering
  \subfloat[Improper alignment.]{\includegraphics[width=0.6\linewidth]{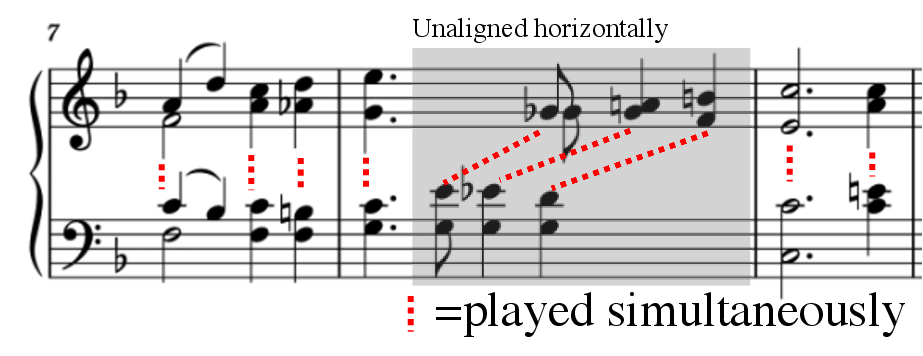}\label{fig:improper-align}}

  %proposed-lh
  \subfloat[Unnatural voice crossing.]{\includegraphics[width=0.45\linewidth]{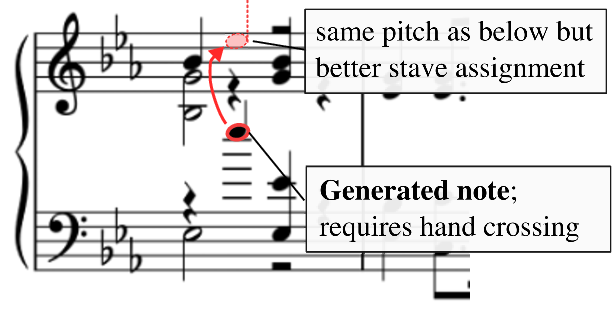}\label{fig:proposed-lh}}
%  \subfloat[Unnatural voice crossing.]{\includegraphics[width=0.6\linewidth]{svg-maezawa/fig-voice-2.eps}\label{fig:oracle-voice-crossing}}
    \subfloat[Unnatural harmony]{
\includegraphics[width=0.23\linewidth]{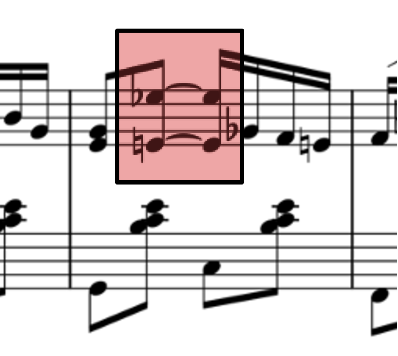}
  \label{fig:proposed-interval}}

  \caption{Some unnatural scores commented by the pianists.}
  \label{fig:score-example}
\end{figure}

The participants left comments about parts that were difficult to read or play.  
The comments were written in a non-English language and translated into English in the subsequent discussions.
In general, for each piece, the pianists tended to make similar remarks about specific measures rather than expressing concerns about the overall structure or musicality of the entire piece.  This suggests that each composition contained a few markedly problematic points, rather than being uniformly unnatural or unplayable throughout.
We show some of the commonly mentioned spots in Figure~\ref{fig:score-example}.

The comments showed that the notation quality is a significant obstacle to sight-readability for both the oracle and the proposed method. 
A common and significant reported problem was the improper horizontal alignment, which increases cognitive load during sight-reading. Pianists commented that 
"\textit{overall, the bars are narrow and somewhat hard to read} (proposed)," 
 "\textit{the vertical are not aligned, making it hard to read} (oracle, Fig.~\ref{fig:improper-align})," 
  and "\textit{ the beats of right and left hands are not aligned} (oracle)." 
Furthermore, the voicing, the assignment of a note to one of the two staves of the piano score, and its beam direction were another concern across both methods. Pianists noted that "\textit{left hand overlaps unnaturally with the right hand's registers} (oracle),"
 "\textit{I get the impression that voice crossing was not intentionally considered} (oracle)," and "\textit{having to play this note on left hand is strange} (proposed, see Fig.~\ref{fig:proposed-lh})."
These observations and the ratings suggest that the generated scores had some parts with confusing notation. 

In addition to readability, the participants reported some musical and playability challenges. For example, the oracle included passages that "\textit{have large leaps in a short interval}," making them difficult to execute. The proposed method also contained passages that require the participant "\textit{to look at [the] hand because the next note jumps too far}" or require "\textit{unnatural hand shape [that] makes mistakes likely.}"
In addition, the proposed method occasionally contained musically unnatural elements, such as parts whose "\textit{chord progression didn't feel right at first}," or contained note combinations that "\textit{seem too harsh for this kind of piece.} (Fig.~\ref{fig:proposed-interval})"

These suggest that the proposed method generated natural and playable music except for some unnatural leaps or pitch intervals.  More importantly, confusing notation confused the participants regardless of whether the data was generated or written by a human.

\section{Ethical Statement}

This work proposes a system to support piano teachers in creating personalized sight-reading exercises with controllable difficulty. 
The goal is twofold: to assist educators by automating repetitive tasks, such as generating varied, level-appropriate material, and to enable students to practice independently with exercises tailored to their skill level.

Although one might be concerned that such a system could negatively affect music teaching jobs, 
we believe that addressing the proposed task does not threaten the role of educators
whose responsibilities extend well beyond the compilation of sight-reading exercises. 
Rather, this technology should be viewed as a supportive tool that helps teachers in their practice.

All training data used in this work come from open and copyright-free sources to ensure compliance with intellectual property rights. 
We also release the code, trained models, and evaluation resources to promote transparency and reproducibility, contributing to open science and enabling further research on personalized music education and difficulty-conditioned score generation.

\section{Conclusion}

This paper presents a method for generating piano exercises with controllable difficulty levels, targeting sight-reading practice. 
By introducing an auxiliary objective for difficulty prediction, we make the model better in following conditioning signals, enabling for more precise and adaptive score generation. 
Although properties such as playability, readability, and completeness were not explicitly supervised, our evaluation suggests that these qualities emerge indirectly from the data and the design of the learning objective.

Although the use of synthetic labels may introduce some degree of noise, collecting a sufficiently large set of expert annotations for generative training is logistically and financially prohibitive. Nevertheless, our results demonstrate that scalable and reliable conditioning can be achieved by leveraging interpretable pedagogical descriptors combined with lightweight classification models.

Although full student personalization remains an open challenge, generative approaches represent an essential step toward this goal, going beyond the limitations of recommendation-based systems. 
Because real sight-reading exams use unseen pieces, generation is particularly well suited to this setting. 
Our method currently models general difficulty trends based on expert annotations from Henle Verlag, but deeper personalization remains a promising direction for future work. 
For example, we envision integrating user feedback loops or personalized difficulty models to better adapt exercises to individual learners.

The system is designed to support, not replace, music educators. 
It allows interactive control at generation time by applying direct constraints during decoding (e.g. key, rhythmic figures, melodic range) without retraining, enabling teachers to adapt exercises to specific pedagogical needs. 
However, complete inpainting would require specific training. 
Collaborations with educators will be explored in future work to investigate how this system could be effectively integrated into classroom practice.

In general, this work contributes to the broader vision of using artificial intelligence to support music education. 
Rather than replacing teachers, our aim is to provide tools that help them tailor practice materials, discover appropriate repertoire, and reduce preparation time. 
We hope that this research encourages further exploration of responsible, pedagogically grounded generative systems for music learning.

\section*{Acknowledgments}
We would like to thank Vsevolod Eremenko for his persistent encouragement over the years to use a Gaussian Naive Bayes classifier for difficulty prediction; in this work, its speed and ability to extract confidence estimates have proven particularly useful.
Pedro also wishes to thank Yamaha Corporation and the Yamaha Global Research Internship program for the opportunity to spend three months in Japan collaborating with outstanding researchers on a project not directly tied to any specific product. We want to thank IMPA Project PID2023-152250OB-I00 funded by MCIU/AEI/10.13039/501100011033/FEDER, UE for partially funding this project.

% ==== CRediT authorship contribution statement ====
\section*{Credit authorship contribution statement}
Pedro Ramoneda: Data Collection, Experiments, Methodology, Writing – original draft.\\  
Masahiro Suzuki: Data Collection, Experiments, Methodology, Writing – review \& editing.\\ 
Akira Maezawa: Methodology, Experiments, Subjective Study, Validation, Writing – review \& editing.\\ 
Xavier Serra: Supervision.

% ==== Declaration of competing interest ====
\section*{Declaration of competing interest}
Part of this research was conducted during a Yamaha Global Research Internship. However, the study is not related to any Yamaha product, and the authors declare no competing financial or personal interests.

\section*{Declaration of generative AI and AI-assisted technologies in the writing process}

During the preparation of this work the authors used ChatGPT (OpenAI) to improve the readability and clarity of the manuscript. After using this tool, the authors carefully reviewed and edited the content to ensure accuracy and appropriateness, and they take full responsibility for the content of the published article.

% ==== Bibliografía ====
\bibliographystyle{elsarticle-num}  % Estilo oficial para KBS
\bibliography{references}          % Cambia "references" por el nombre real de tu .bib

% ==== Apéndices (opcional) ====
% \appendix
% \section{Additional experiments}
% Aquí puedes incluir material complementario si es necesario.

\end{document}